\documentclass[twocolumn,reprint,aps,prd,nofootinbib,floatfix,superscriptaddress]{revtex4-1}
\usepackage[utf8]{inputenc}
\usepackage[utf8]{inputenc}
\usepackage{amsmath}
\usepackage{amsfonts}
\usepackage{amssymb}
\usepackage[left=2cm,right=2cm,top=2cm,bottom=2cm]{geometry}
\usepackage{braket}
\usepackage{dsfont}
\usepackage{hyperref}

\usepackage{graphicx}
\usepackage{tikz}
\usetikzlibrary{arrows,decorations.pathmorphing,backgrounds,positioning,fit,petri,shapes}
\usepackage{lipsum}

\usepackage{multirow}
\begin{document}

\title{Cosmic Shear: Inference from Forward Models}

\begin{abstract}
 Density-estimation likelihood-free inference (DELFI) has recently been proposed as an efficient method for simulation-based cosmological parameter inference. Compared to the standard likelihood-based Markov Chain Monte Carlo (MCMC) approach, DELFI has several advantages: it is highly parallelizable, there is no need to assume a possibly incorrect functional form for the likelihood and complicated effects (e.g the mask and detector systematics) are easier to handle with forward models. In light of this, we present two DELFI pipelines to perform weak lensing parameter inference with lognormal realizations of the tomographic shear field -- using the $C_\ell$ summary statistic. The first pipeline accounts for the non-Gaussianities of the shear field, intrinsic alignments and photometric-redshift error. We validate that it is accurate enough for Stage III experiments and estimate that $\mathcal{O}(1000)$ simulations are needed to perform inference on Stage IV data. By comparing the second DELFI pipeline, which makes no assumption about the functional form of the likelihood, with the standard MCMC approach, which assumes a Gaussian likelihood, we test the impact of the Gaussian likelihood approximation in the MCMC analysis. We find it has a negligible impact on Stage IV parameter constraints. Our pipeline is a step towards seamlessly propagating all data-processing, instrumental, theoretical and astrophysical systematics through to the final parameter constraints.
\end{abstract}

\author{Peter L.~Taylor}
\email{peterllewelyntaylor@gmail.com}
\affiliation{Mullard Space Science Laboratory, University College London, Holmbury St.~Mary, Dorking, Surrey RH5 6NT, UK}
\author{Thomas D.~Kitching}
\affiliation{Mullard Space Science Laboratory, University College London, Holmbury St.~Mary, Dorking, Surrey RH5 6NT, UK}
\author{Justing~Alsing}
\affiliation{Center for Computational Astrophysics, Flatiron Institute, 162 5th Ave, New York City, NY 10010, USA}
\affiliation{Oskar Klein Centre for Cosmoparticle Physics, Stockholm University, AlbaNova, Stockholm SE-106 91, Sweden}
\affiliation{Imperial Centre for Inference and Cosmology, Department of Physics, Imperial College London, Blackett Laboratory, Prince Consort Road, London SW7 2AZ, UK}
\author{Benjamin D. Wandelt}
\affiliation{Center for Computational Astrophysics, Flatiron Institute, 162 5th Ave, New York City, NY 10010, USA}
\affiliation{Sorbonne Universit\'e, CNRS, UMR 7095, Institut d’Astrophysique de Paris, 98 bis boulevard Arago, 75014 Paris, France}
\affiliation{Sorbonne Universit\'e, Institut Lagrange de Paris (ILP), 98 bis boulevard Arago, 75014 Paris, France}
\affiliation{Department of Astrophysical Sciences, Princeton University, Princeton, NJ 08540, USA}
\author{Stephen M. Feeney}
\affiliation{Center for Computational Astrophysics, Flatiron Institute, 162 5th Ave, New York City, NY 10010, USA}
\author{Jason D.~McEwen}
\affiliation{Mullard Space Science Laboratory, University College London, Holmbury St.~Mary, Dorking, Surrey RH5 6NT, UK}
%\date{23 November 2018}
\maketitle

\section{Introduction}
Weak lensing by large scale structure offers some of the tightest constraints on cosmological parameters. Over the next decade data from Stage IV experiments including Euclid\footnote{\url{http://euclid-ec.org}} \cite{laureijs2010euclid}, WFIRST\footnote{\url{https://www.nasa.gov/wfirst}} \cite{spergel2015wide} and LSST\footnote{\url{https://www.lsst.org}} \cite{anthony4836large} will begin taking data. Extracting as much information from these ground-breaking data sets, in an unbiased way, presents a formidable challenge. 
\par  The majority of cosmic shear studies to date focus on extracting information from two-point statistics and in particular the correlation function, $\xi (\theta)$, in configuration space and the lensing power spectrum, $C_\ell$, in spherical harmonic space~\cite{heymans2013cfhtlens, troxel2017dark, kitching20143d, hildebrandt2017kids, hikage2018cosmology}. While the non-Gaussian information in the shear field is accessed with higher-order statistics~\cite{semboloni2010weak, fu2014cfhtlens}, peak counts~\cite{peel2017cosmological, jain2000statistics} or machine learning~\cite{gupta2018non}, the impact of systematics on the two-point functions have been extensively studied~\cite{massey2012origins}. For this reason we will focus on these statistics and leave the higher-order information to a future study. In particular we focus on the $C_\ell$ statistic because computing correlation functions from catalogues with billions galaxies -- even using an efficient code such as {\tt TREECORR}~\cite{jarvis2004m} -- is extremely computationally demanding. 
\par Apart from~\cite{alsing2015hierarchical,alsing2016cosmological}, existing studies of the shear two-point statistics~\cite{heymans2013cfhtlens, troxel2017dark, kitching20143d, hildebrandt2017kids, hikage2018cosmology} use a Gaussian likelihood analysis to infer the cosmological parameters. This approach has drawbacks. For example, with the improved statistical precision of next generation data, we will need to propagate complicated `theoretical systematics' (e.g. reduced shear~\cite{dodelson2006reduced}) and detector effects~\cite{massey2012origins} into the final cosmological constraints. It is difficult to derive the expected impact of these effects as is required for a likelihood analysis. It is much easier to produce forward model realizations. 
\par It has also recently been claimed that because the true lensing likelihood is left-skewed, not Gaussian, parameter constraints from correlation functions are biased low in the $\sigma_8 - \Omega_m$ plane~\cite{sellentin2017insufficiency,sellentin2018skewed}. The same argument given in these papers applies to the $C_\ell$ statistic. More will be said about this in Section~\ref{sec:Inference}.
\par To overcome these issues, a new method called density-estimation likelihood-free inference (DELFI)~\cite{bonassi2011bayesian, papamakarios2018sequential, fan2013approximate, papamakarios2016advances, alsing2018massive, lueckmann2018likelihood, alsingndes} offers a way forward. DELFI is a `likelihood-free' method similar to approximate Bayesian computation (ABC)~\cite{ABC}, but much more computationally efficient. Using summary statistics (the $C_\ell$ statistic, in this case) generated from full forward models of the data at different points in cosmological parameter space, DELFI is used to estimate the posterior distribution.
\par Performing inference on realizations of the data may seem computationally challenging, but using efficient data compression~\cite{alsing2018massive, alsing2018generalized} most applications require only $\mathcal{O} (1000)$ simulations~\cite{alsingndes}. This is less than the number of simulations already required to produce a valid estimate of the inverse covariance matrix in a Stage IV likelihood analysis. DELFI is also highly parallelizable. 
\par Two additional likelihood-free methods are introduced in~\cite{leclercq2018bayesian, leclercq2019primordial}. The aims of these methods respectively are to optimally choose points in parameter space, reducing the number of simulations, and to infer a larger number of parameters. Nevertheless we choose to work with DELFI because cosmic shear simulations are expensive, so we want to take advantage of parallelism and we only need to infer a small number of cosmological parameters. Methods to deal with a large number of nuisance parameters in DELFI are discussed in~\cite{alsingnuissance}.
\par
\begin{figure}
\centering
\includegraphics[width = 80mm]{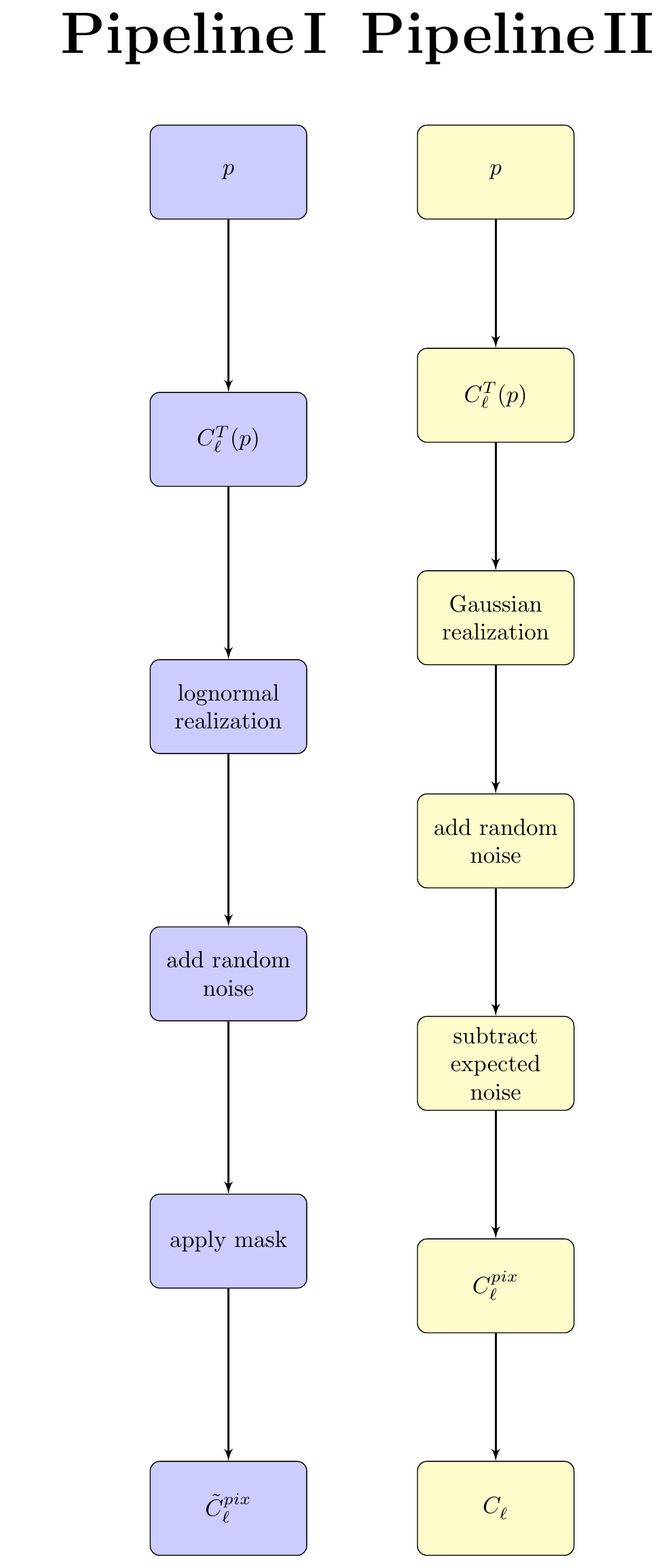}
\caption{A schematic of the two forward model pipelines used in this work given model parameters $p$. In Pipeline I we develop a forward model of cosmic shear data for inference with DELFI which takes advantage of the forward model approach. There is no need to deconvolve the mask or pixel window function, for example. In Pipeline II we use a Gaussian field, do not use a mask, subtract off the shot-noise or deconvolve the pixel window function. These choices allow us to make a direct comparison between DELFI and a Gaussian likelihood analysis to test the Gaussian likelihood assumption.}
\label{fig:pipeline}
\end{figure}

 The goals of this paper are threefold:
\begin{itemize}
\item To develop a more realist forward model of the shear field than the one presented in~\cite{alsingndes}, including the impact of intrinsic alignments and non-Gaussinities of the field, and determine whether this changes the number of simulations needed to perform inference with DELFI.
\item To test the impact of the Gaussian-likelihood assumption used in nearly all cosmic shear studies and in so doing test whether the data compression of the $C_\ell$ summary statistic used in DELFI is lossless.
\item To validate that the forward model presented in this paper will be accurate enough to perform inference on today's stage III data sets.
\end{itemize}
To achieve these aims we develop two cosmic shear forward model pipelines for DELFI, using the publicly available {\tt pydelfi}\footnote{\url{https://github.com/justinalsing/pydelfi/commits/master}} implementation, summarized in Figure~\ref{fig:pipeline}. Pipeline I takes full advantage of the benefits of forward modelling and is intended for application to real data-sets, while Pipeline II is intended only for comparison with the standard likelihood analysis. We also consider 3 different analyses summarized in Table~\ref{tab1}. DA1 is a DELFI analysis using shear Pipeline I. Meanwhile we compare the DELFI analysis, DA2, to the likelihood analysis, LA, to test the impact of the Gaussian likelihood approximation. It is useful for the reader to refer back to Table~\ref{tab1} and Figure~\ref{fig:pipeline} throughout the text. 
\par The structure of this paper is as follows. The formalism of cosmic shear and cosmological parameter inference is reviewed in Sections~\ref{sec: cosmic shear formalism}-\ref{sec:Inference}. While DELFI has already been applied to cosmic shear in a simple Gaussian field setting~\cite{alsingndes}, in Section~\ref{sec:forward model} we go beyond this and present a more realistic forward model (Pipeline I) which includes the impact of intrinsic alignments and non-Gaussianities of the shear field. We also estimate the number of simulations required for a Stage IV experiment and check to confirm that we recover the input cosmology from a DA1 anlysis on mock data. Next we discuss the feasibility of the DA1 analysis for Stage III data in Section~\ref{sec:stage iii}. In Section~\ref{sec:test gauss like} we test the impact of the Gaussian likelihood approximation by comparing the DA2 analysis to the LA analysis. In Section~\ref{sec:Future } we discuss future prospects for DELFI in cosmic shear studies, before concluding in Section~\ref{sec:Conslusion}.

\begin{table*}
\begin{tabular}{ |p{4cm}||p{4cm}|p{4cm}|p{4cm}|  }
 \hline
 \hline 
 \text{ } & DA1 & DA2 & LA \\
 \hline
 Inference   & DELFI    & DELFI &   Gaussian likelihood\\
  Pipeline   & Pipeline I & Pipeline II &  NA\\
 Number of galaxies &   $1.56 \times 10 ^9$  & $1.56 \times 10 ^9$   & $1.56 \times 10 ^9$\\
 Number of tomographic bins &6 & 2 &  2\\
 Number of $\ell$-bins &15 with $\ell \in [10,1000]$ & 15 with $\ell \in [10,1000]$ &  15 with $\ell \in [10,1000]$\\
 Field type    & Lognormal & Gaussian &  NA\\
Deconvolve Pixel Window &   No  &  Yes & NA\\
 Mask & Yes  & No   & NA\\
 Subtract shot-noise & No  & Yes & NA\\
 \hline
\end{tabular}
\caption{The three analyses in this paper. In DA1 we use DELFI to infer the cosmological parameters. Since we perform inference with forward models, there is no need to deconvolve the pixel window function, deconvolve the mask or subtract off the shot noise. This analysis is applied to mock Stage IV data in Section~\ref{sec:forward model}. Meanwhile by comparing the DELFI analysis, DA2, with the likelihood analysis, LA, we test the impact of the Gaussian likelihood approximation. In DA2 our modeling choices are governed by the constraint that we must match the Gaussian likelihood analysis as closely as possible. Some of the map-level choices are not applicable to the likelihood analysis LA.}
\label{tab1}
\end{table*}

\section{Cosmic Shear Formalism and the Lognormal Field Approximation} \label{sec: cosmic shear formalism}
\subsection{The Lensing Spectrum} \label{sec:conv}

\par Assuming the Limber~\cite{loverdelimber, kitchinglimits}, spatially-flat Universe~\cite{PLTtesting}, flat sky~\cite{kitchinglimits} and equal-time correlator approximations~\cite{kitchingunequal}, the lensing spectrum, $C_{\ell,\rm GG}^{ij}$, is given by~\cite{heymans2013cfhtlens}:
\begin{equation}
C_{\ell, \rm GG}^{ij} = \int_0^{r_{\rm H}} dr \, 
\frac{q_i(r)q_j(r)}{r^2} \, P \left( \frac{\ell}{r},r \right),
\label{eqn:CGG} 
\end{equation}
where $P(k,r)$ is the matter power spectrum and the lensing efficiency kernel, $q_i$ is defined as:
\begin{equation}
q_i(r) = \frac{3 H_0^2 \Omega_{\rm m}}{2c^2} \frac{r}{a(r)}\int_r^{r_{\rm H}}\, dr'\ n_i(r') 
\frac{r'-r}{r'}, 
\label{eqn:qk} 
\end{equation}
and we generate the $N_{\text{tomo}}$ tomographic bins, $n_i(r')$, by dividing the radial distribution function:
\begin{equation} \label{eq:n(z)}
n \left( z_p \right) \propto \frac{a_1}{c_1} {\rm e} ^ {- \frac{ \left( z-0.7 \right) ^2 }{b_1^2} } + {\rm e} ^ {- \frac{ \left( z-1.2 \right) ^2 }{d_1^2} } ,
\end{equation}
with $\left(a_1/c_1,b_1 ,d_1 \right) =\left( 1.5 / 0.2, 0.32, 0.46 \right)$ \cite{van2013cfhtlens} into bins with an equal number of galaxies per bin.
To account for photometric redshift error, each bin is smoothed by the Gaussian kernel:
\begin{equation} \label{eq:photo error}
p \left( z | z_p \right) \equiv \frac{1}{2 \pi \sigma_z \left(z_p \right)} {\rm e} ^{- \frac{ \left( z -c_{\rm cal} z_p + z_{\rm bias} \right) ^2  } {2 \sigma_{z_p}} },
\end{equation}
with $c_{\rm cal} = 1$, $z_{\rm bias} = 0$ and $\sigma_{z_p} = A \left(  1 + z_p \right)$ with $ A = 0.05$ \cite{ilbert2006accurate}.

\subsection{Intrinsic Alignments} \label{sec:IA}
The tidal alignment of galaxies around massive halos adds two additional terms to the lensing spectrum. An `II term' accounts for the intrinsic tidal alignment of galaxies around massive dark matter halos, while a `GI term' accounts for the anti-correlation between tidally aligned galaxies at low redshifts and weakly lensed galaxies at high redshift.
\par We model this effect using the non-linear alignment (NLA) model~\cite{hirata2004intrinsic, heymans2013cfhtlens}. We also allow the intrinsic amplitude, $A(z)$, to vary as a function of redshift so that $A(z) = \left[ ( 1+ z_0) / (1+z) \right] ^ \eta $~\cite{maccrann2015cosmic}, where $z_0$ is the mean redshift of the survey. This is $z_{0} = 0.76$ for the $n(z)$ given in equation~(\ref{eq:n(z)}). This model was used in the joint KiDS-450+2dFLenS~\cite{joudaki2017kids} analysis and was one of the models considered in the Dark Energy Survey Year 1 cosmic shear analysis (hereafter DESY1)~\cite{troxel2017dark}.
\par In this case the II spectrum, $C_{\ell, \rm II}^{ij}$, is given by:
\begin{equation}
C_{\ell,\rm II}^{ij} = \int_0^{r_{\rm H}} dr \, 
\frac{n_i(r)n_j(r)}{r^2} \, P_{\rm II} \left( \frac{\ell}{r},r \right),
\label{eqn:CII} 
\end{equation}
where the II matter power spectrum is:
\begin{equation} \label{eqn:PII}
P_{\rm II}(k,z) =  F^2(z) P(k,z)
\end{equation}
and
\begin{equation} \label{eqn:Fz} 
F(z) = - A_I C_1 \rho_{\rm crit} \frac{\Omega_{\rm m}}{D(z)} \left( \frac{1 + z \left( r\right) }{1 + z_0}\right) ^{\eta} ,
\end{equation}
where $\rho_{\rm crit}$ is the critical density of the Universe, $D(z)$ is the growth factor and $C_1 = 5 \times 10^{-14} h^{-2} M_\odot^{-1} {\rm Mpc}^3$. 
\par The GI power spectrum is:
\begin{equation}
C_{\ell, \rm GI}^{ij} = \int_0^{r_{\rm H}} dr \, 
\frac{q_i(r)n_j(r) + n_i(r)q_j(r) }{r^2} \, P_{\rm GI} \left( \frac{\ell}{r},r \right),
\label{eqn:CGI} 
\end{equation}
and the GI matter power spectrum is:
\begin{equation} \label{eqn:PGI}
 P_{\rm GI}(k,z) =  F(z) P (k,z).
\end{equation}
Altogether the theoretical lensing spectrum, $C^{T, ij}_\ell$, is given by the sum of the three contributions:
\begin{equation}
    C^{T}_\ell = C_{\ell, \rm GG}^{ij} + C_{\ell, \rm GI}^{ij} + C_{\ell, \rm II}^{ij}
\end{equation}
Henceforth we will routinely drop the tomographic bin labels for convenience, as we have done here, on the left hand side.
\par All lensing spectra are generated inside the {\tt Cosmosis} framework~\cite{cosmosis}. The linear power spectrum and expansion history are computed with {\tt CAMB}~\cite{camb} and the nonlinear corrections are computed with {\tt HALOFIT}~\cite{halofit}.

\subsection{The Lognormal Field Approximation} \label{sec:lognormal}
\begin{figure}
\centering
\includegraphics[width = 85mm]{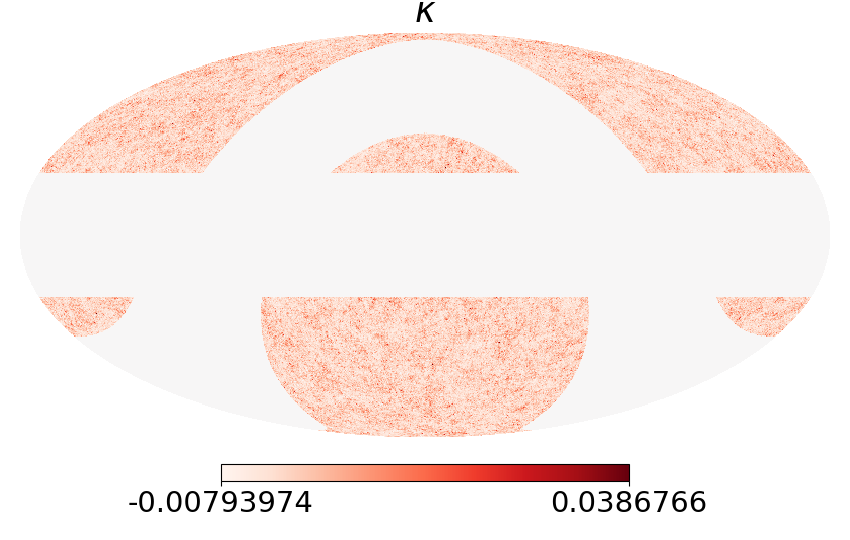}
\includegraphics[width = 85mm]{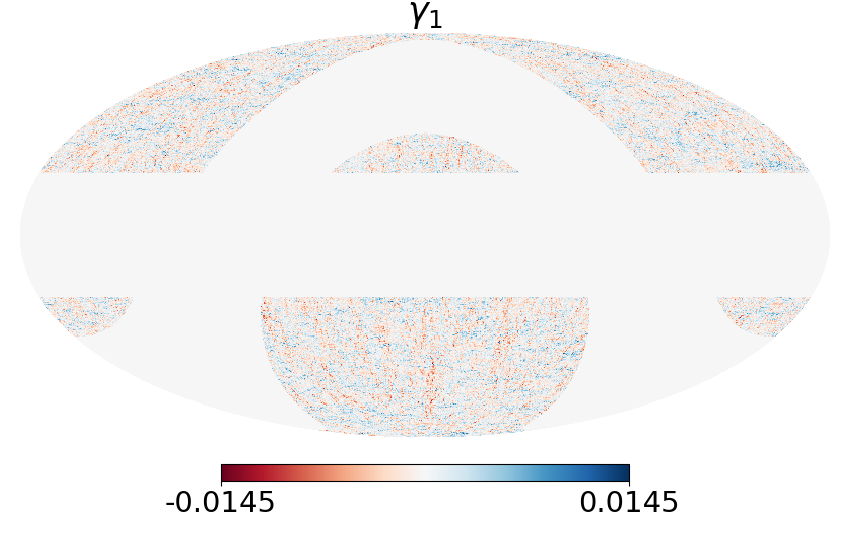}
\includegraphics[width = 85mm]{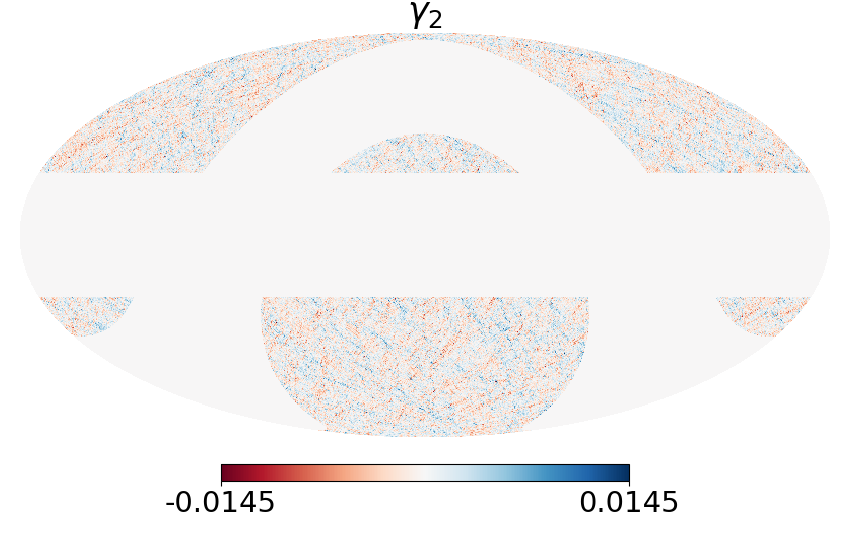}
\caption{A single masked data realization of the convergence field, $\kappa$, and the two observable shear components: $\gamma_1$ and $\gamma_2$ (including shape noise) for a typical Stage IV experiment.  This is the lowest redshift bin of six, where the effect of non-Gaussianity is largest. The non-Gaussianity is clearly visible in the $\kappa$-map (the color scale runs between the minimum and maximum value of the $\kappa$-field to make the non-Gaussianity more visible), where the majority of pixels are very slightly negative with a small number of pixels taking very large (positive) $\kappa$-values. The mask cuts all pixels lying within $22.5 \text{ deg}$ of the galactic and ecliptic planes.}
\label{fig:fields}
\end{figure}

Generating lognormal convergence fields~\cite{lognormalschneider} is computationally inexpensive and captures the impact of nonlinear structure growth more accurately than Gaussian realizations. This approximation was recently used in DESY1~\cite{troxel2017dark} to compute the covariance matrix from noisy realisations of the data. No differences in parameter constraints were found when the covariance was computed using lognormal fields compared to the halo model approach~\cite{krause2017cosmolike}.
\par In the lognormal field approximation the convergence, $\kappa ^ i  (\theta)$, inside each tomographic bin, $i$, is generated by exponentiating and shifting a Gaussian realization, $g ^i \left( \theta \right)$, according to:
\begin{equation} \label{eq:lognorm}
    \kappa ^ i  (\theta) = \text{exp} \left[g ^i \left( \theta \right) \right] - \kappa^i_0,
\end{equation}
where $\kappa^i_0$ is a constant shift parameter.
\par We use {\tt Flask}~\cite{flask} to generate consistent lognormal realizations~\cite{lognormalschneider} of the convergence and shear fields -- correlated between redshift slices. The procedure is discussed in detail in Section 5.2 of~\cite{flask} (see also~\cite{mancini20183d}). 
\par {\tt Flask} takes just two inputs: 
\begin{itemize}
\item {\tt Flask} takes the theoretical lensing spectrum, $C^T_\ell$, defined in Sections~\ref{sec:conv}-\ref{sec:IA}. Formally {\tt Flask} uses the convergence spectrum to generate a convergence field, $\kappa$, from which it computes a consistent shear field, $\gamma$. In the flat sky approximation -- which we assume throughout -- the shear and convergence spectrum are the same, but care would be needed to correctly re-scale the input convergence spectrum by the appropriate $\ell$-factor if the flat sky approximation was dropped~\cite{kitchinglimits, castro}.
\item {\tt Flask} requires the shift parameter, $\kappa^i_0$ for each tomographic bin $i$. We compute this by taking a weighted average of the shift parameter at each redshift:
\begin{equation}
    \kappa^i_0 = \int \text{d}z \text{ }  n_i(z) \kappa^i_0 (z),
\end{equation}
using the fitting formula:
\begin{equation}
    \kappa^i_0 (z) = 0.008z + 0.029 z^2 - 0.0079 z^3 + 0.0065 z^4
\end{equation}
derived from simulations~\cite{lognormalschneider}. 
\end{itemize}
\par While the fitting formula will have some cosmological dependence, the shift parameter does not affect the  power spectrum of the field -- only impacting cosmological constraints through the covariance. Non-Gaussian corrections to the covariance already have a sub-dominant impact~\cite{sato2013impact,eifler2014combining}, hence the dependence of these corrections on the cosmology is further sub-dominant. For this reason we ignore the cosmological dependence of the shift parameter.
\par A valid covariance matrix between data must be positive-definite, but this is not guaranteed for correlations between tomographic lognormal fields~\cite{flask}. {\tt Flask} overcomes this issue by perturbing the lognormal fields following the regularization procedure outlined in Section 3.1 of~\cite{flask}. Provided that the regularization is applied to a small number of tomographic bins, it is found in~\cite{flask} that $C_\ell^{\text{reg}} / C_\ell^{{\rm ln}} \ll 1 \times 10 ^{-5}$, where $C_\ell^{\text{reg}}$ is the recovered regularized spectrum and $C_\ell^{{\rm ln}}$ is the spectrum recovered from the unregularized map~\cite{flask}. In Section~\ref{sec:stage iii} we verify that this will not impact Stage III parameter constraints.
\par In Figure~\ref{fig:fields} we plot a single lognormal realization generated with Pipeline I. We show the masked convergence and components of the shear field in the lowest redshift bin. This is where the non-Gaussianities are most pronounced and clearly visible. In the convergence map the majority of the pixels take small negative values. However there are rare incidences of large positive convergence. This physically corresponds to collapsed high-density structures along the line-of-sight.

\subsection{Band-limit Bias from the Lognormal Field} \label{sec:band-limit}
\par Unlike Gaussian fields, lognormal realizations are not band-limited in $\ell$~\cite{flask} (see Section 5.2.2 therein). In particular, Taylor expanding the lognormal convergence field, $\kappa ^ i  (\theta)$, in terms of the Gaussian field, $g ^i \left( \theta \right)$, yields quadratic and higher order terms in $g^i$. In harmonic space this mixes different $\ell$-modes. When a band-limit is imposed, this biases the lensing spectrum recovered from the map.

\section{Cosmological Parameter Inference} \label{sec:Inference}
\subsection{Gaussian likelihood Analysis} 
In the standard two-point cosmic shear likelihood analysis, we assume a Gaussian likelihood:
\begin{equation} \label{eq:gauss}
\text{ln } \mathcal{L}\left( p\right) = - \frac{1}{2} \sum_{a,b} \left[ D_a - T_a\left( p \right) \right] C_{ab}^{-1} \left[ D_b - T_b \left( p\right) \right],
\end{equation}
where $ D_a$ and $T_a\left( p \right)$ are the data and theory vectors respectively composed of the $C_\ell$ estimated from data and the theoretical expectation of $C_\ell$ given cosmological parameters $p$. 
\par Meanwhile $C_{ab}^{-1}$ is the inverse of the covariance matrix. Since we generate the covariance matrix from noisy simulations of the data we make the Anderson-Hartlap~\cite{hartlap2007your, anderson1958introduction} correction, to avoid bias from inverting the covariance matrix, for the remainder of the paper.

\subsection{The Potential Insufficiency of the Gaussian likelihood Approximation in Cosmic Shear} \label{sec:insufficient}

\begin{figure}
\centering
\includegraphics[width = 68mm]{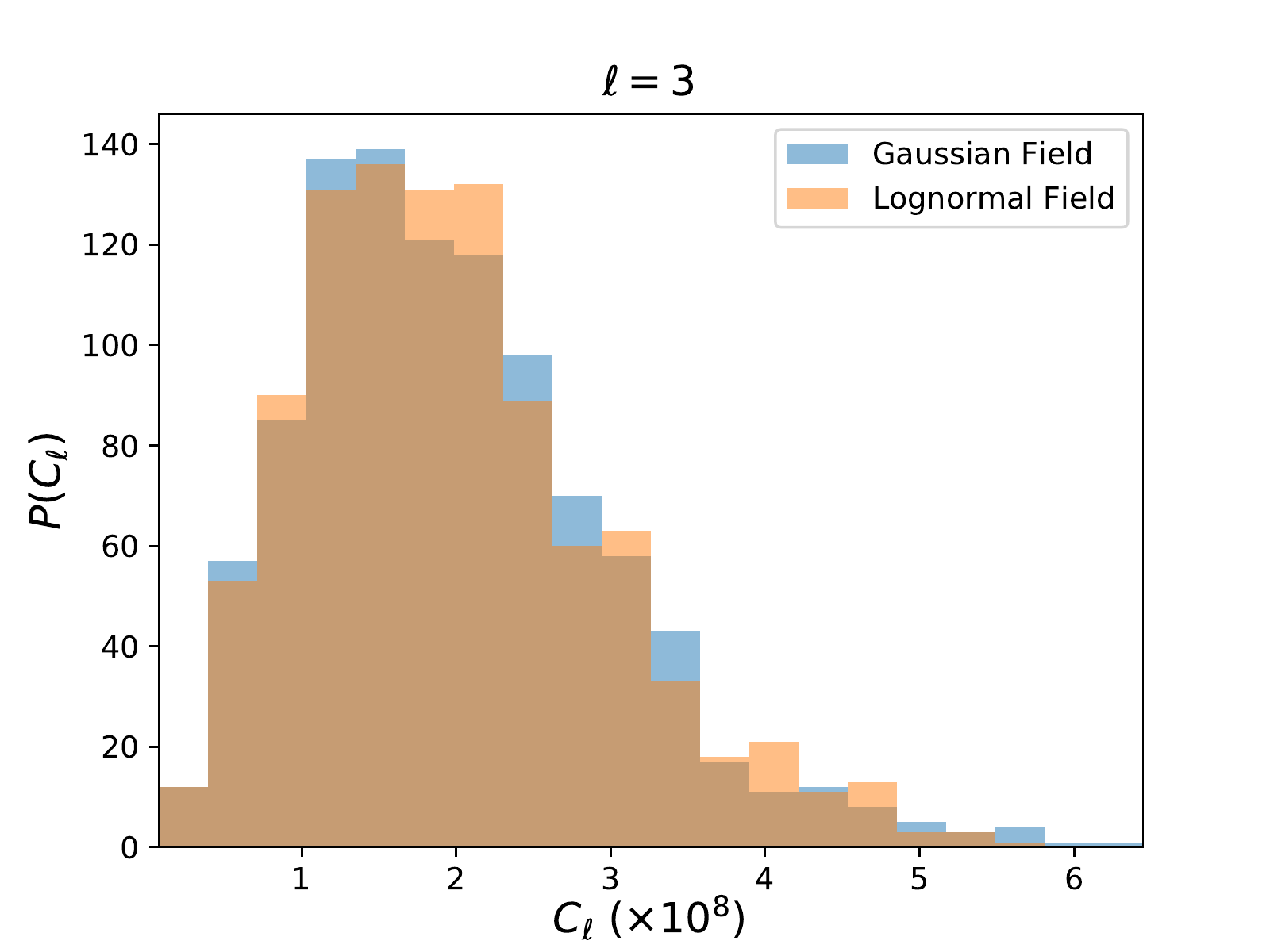}
\includegraphics[width = 68mm]{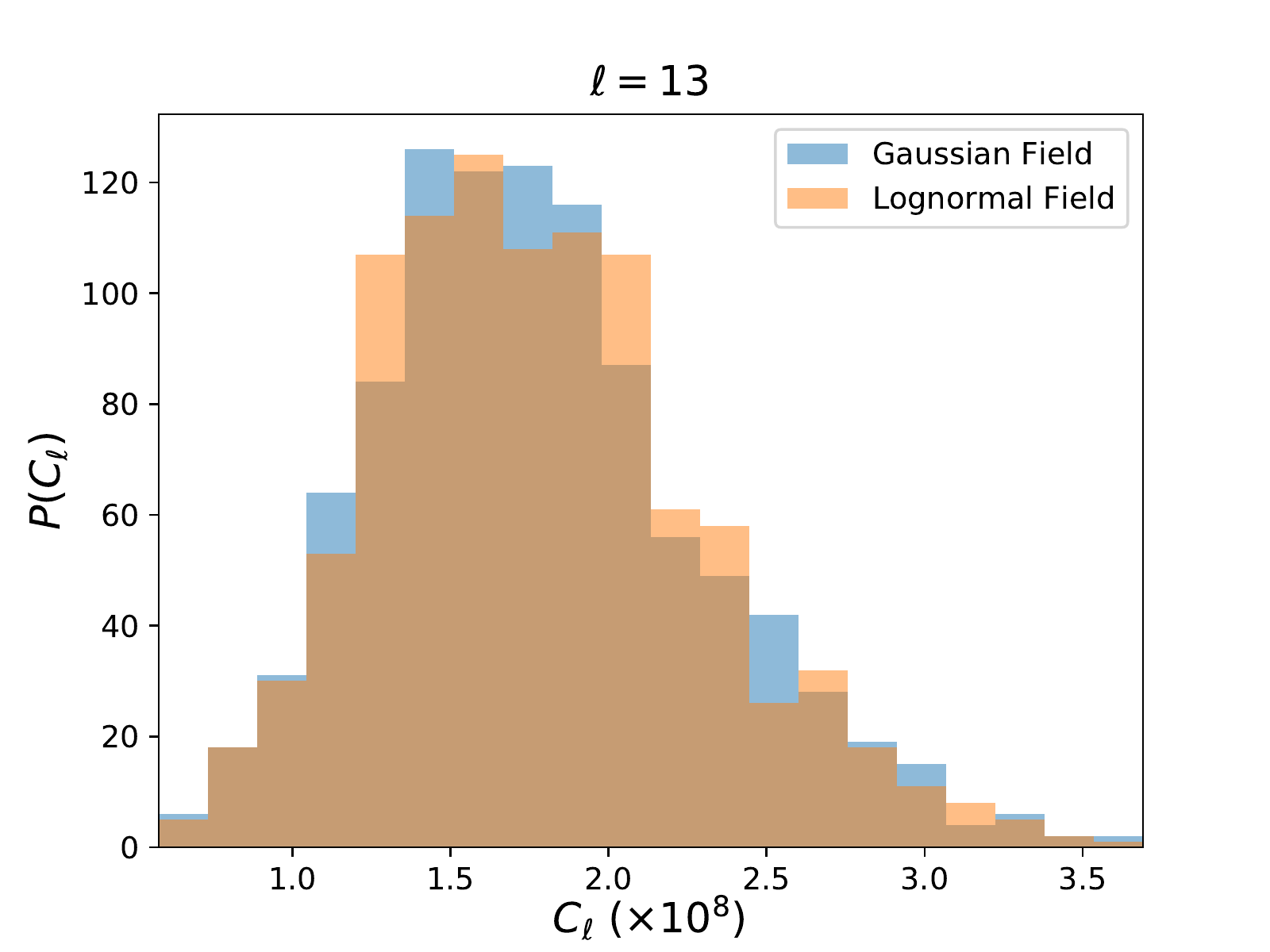}
\includegraphics[width = 68mm]{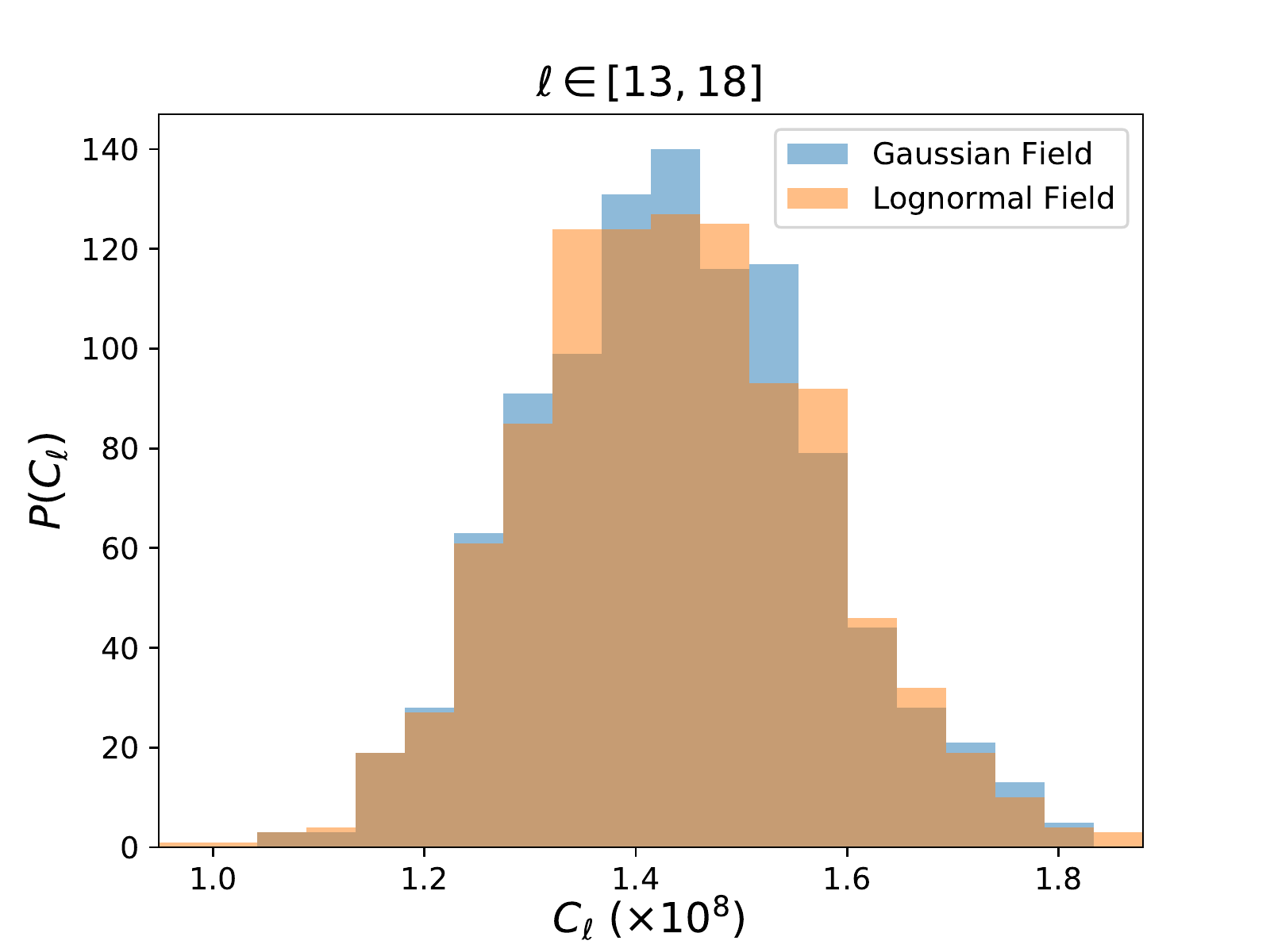}
\caption{The distribution of $C_\ell$ drawn from 1000 Gaussian (blue) and lognormal (orange) realizations. Each subplot corresponds to different $\ell$-mode and binning strategy. In all cases there is little difference in the skew of the distribution when using Gaussian or lognormal fields. This suggests the impact of the Gaussian likelihood approximation can be tested under the assumption of a Gaussian field, as in this work. {\bf Top:} The $\ell = 3$ case is clearly skewed, as expected for low-$\ell$ (see Section~\ref{sec:insufficient}). {\bf Middle:} The $\ell = 13$ case is slightly less skewed, as expected. {\bf Bottom:} The binned case for $\ell \in [13,15]$.  By the central limit theorem, the distribution is Gaussianized when drawing randomly from the distribution of each independent $\ell$-mode.}
\label{fig:hist}
\end{figure}

To see why the Gaussian likelihood assumptions can lead to bias we summarize the argument given in~\cite{sellentin2018skewed}. Inside a single bin the unmasked lensing spectrum is:
\begin{equation}
    C_\ell = \frac{1}{2 \ell + 1} \sum_{m = - \ell} ^{\ell} \mid \gamma_ {\ell m} \mid ^ 2.
\end{equation}
Since the harmonic coefficients, $\gamma_ {\ell m}$, are computed as a summation over a large number of pixels, they are Gaussian distributed by the central limit theorem. Squaring a Gaussian random variable gives a gamma distribution -- which is left-skewed. This is illustrated in the top two rows of Figure~\ref{fig:hist}.
\par Taking a Gaussian rather than a gamma distribution for the likelihood could bias parameter constraints. Since $S_8^2 =\sigma_8^2 (\Omega_m / 0.3)$ and $\Omega_m$ enter into the shear spectrum amplitude, we would expect these parameters to be ones which are most affected -- and biased low. Only in the limit of large $\ell$ -- as the $C_\ell$ itself becomes the sum over a large number of $m$-modes -- does the central limit theorem kick in and the likelihood become Gaussian. This is illustrated in Figure~\ref{fig:hist} and can be seen by comparing the second and third row.

\subsection{Density-estimation Likelihood-free Compression}
Since density-estimation likelihood-free inference methods are most effective in low dimensions~\cite{alsing2018massive}, we compress the $C_\ell$ summary statistic. As suggested in~\cite{alsing2018generalized}, the lensing spectra are compressed into a new vector, $t$, according to:
\begin{equation} \label{eq:compress1}
    C_\ell \rightarrow t = \nabla _ p \mathcal{L} _\ast,
\end{equation}
where $p$ is the set of cosmological parameters that we are inferring, $\mathcal{L} _\ast$ is a proposal Gaussian likelihood centred at a fiducial set of parameters which we take to be $ \left(\Omega_m, h_0, \Omega_b, n_s, S_8, A, \eta \right) = \left(0.3, 0.72, 0.96,0.79,1.,2.8 \right) $ throughout, where $S_8 = \sigma_8 (\Omega_m /0.3) ^ {0.5}$, $A$ and $\eta$ are the intrinsic alignment parameters defined in Section~\ref{sec:IA} and the other parameters take their standard cosmological definitions.  As the assumption of a Gaussian likelihood here is only for compression purposes, it does not bias the final parameter constraints and the Fisher information is preserved provided the true likelihood is Gaussian~\cite{alsing2018generalized}. If the true likelihood is not exactly Gaussian, some information will be lost. This is investigated in Section~\ref{sec:test gauss like}. For more advanced compression techniques using neural networks, see~\cite{charnock2018automatic}.

\subsection{Density-estimation Likelihood-free Inference} \label{sec:delfi}
We use {\tt pydelfi}~\cite{alsingndes} to learn the conditional density $\mathbb{P} (t |p )$ (this software comes with many different run-mode options, but we restrict our attention to the methods used in this work). The likelihood is then given by $\mathbb{P} (t = t_{\text{data}} |p )$, where $t_{\text{data}}$ is the mock data generated from either Pipeline I or II. Multiplying by the prior, which we take to be flat in all parameters, yields the posterior. 
\par Using the default setting in {\tt pydelfi}, we train five neural density estimators (NDE) (four mixture density networks (MDN) and one masked autoregressive flow (MAF), see~\cite{alsingndes} for more details) with the default network architectures described in Section 4 of~\cite{alsingndes}, parameterized in terms of a set of neural network weights, $ w$. Training multiple networks allows DELFI to avoid over-fitting and increases robustness.
\par We use sequential learning to learn the weights, $w$, updating our knowledge of the conditional density distribution $\mathbb{P} (t = t_{\text{data}}|p )$. Specifically we divide the inference task into 20 training steps with 100 simulations per step. Given a large enough computer all the simulations in each training step could be run in parallel, so that the total time of the simulations would not exceed the time it took to perform 20 simulations.
\par As an initial guess for the conditional distribution, we take the multivariate Gaussian:
\begin{equation}
    \mathbb{P} (t|p) = \mathcal{N} (t| p, F^{-1}),
\end{equation}
where $F^{-1}$ is the inverse of the Fisher matrix of the cosmological parameters, $F = - \langle \nabla_p \nabla_p ^{Tr} \mathcal{L}_* \rangle $. At each step thereafter, we train each neural density estimator on a set of parameter realization pairs $\{p_i, t_i\}$ drawing samples from the conditional density of the previous step to ensure that the highest density regions are the most finely sampled. Meanwhile ten percent of the samples are retained as a validation set to avoid over-fitting. 
\par At each step each NDE learns the weights, $w$, by  minimizing the negative loss function:
\begin{equation} \label{eq:negative loss}
     - \ln U(w) = - \sum_{i = 1}^{N_{\text {samples}}} \ln  \mathbb{P} (t_i |p , w),
\end{equation}
which is an estimate of the Kullback-Leibler divergence between the density estimator $\mathbb{P} (t = t_{\text{data}} |p )$ and the true distribution~\cite{alsingndes} (the minus sign is pulled out of $U(w)$ on the left-hand side following the convention in~\cite{alsingndes}). The final estimate for the conditional distribution is given as a weighted average over the estimates from the five networks:
\begin{equation}
    \mathbb{P} (t| p; w) = \sum_{i\in\text{networks}} \beta_i \mathbb{P}_i(t|p;w),
\end{equation}
where the weights are determined by the relative likelihood of each NDE~\cite{alsingndes}.

\section{The Full Forward Model} \label{sec:forward model}
In this section we use Pipeline I to generate mock Stage IV data and then run analysis DA1 to recover the input cosmology. This allows us to test our pipeline and estimate the number of simulations needed for a Stage IV experiment. We describe the model choices and results below.

\subsection{The Mask} \label{sec:mask}
We use a typical Stage IV survey mask shown in Figure~\ref{fig:fields}. All pixels lying within $22$ deg of either the galactic or ecliptic planes are masked. This leaves 14,490 $\text{deg} ^2$ of unmasked pixels which, as a fraction of the full sky, is $f_{\text{sky}} = 0.35$.

\subsection{Shot-Noise Model}
The noise, $\gamma  _p$, for each pixel, $p$, is drawn from a Gaussian distribution~\cite{alsing2015hierarchical}:
\begin{equation}
\gamma  _p \sim \mathcal{N} \left(0,\frac{\sigma _\epsilon}{\sqrt{\widehat N_P}} \right)
\end{equation}
where $\widehat N_P $ is the number of galaxies in each pixel, the orientation is angle is drawn from a uniform distribution, we take the intrinsic shape dispersion as $\sigma_\epsilon = 0.3$~\cite{brown2002measurement} and use $30 \text{ galaxies per arcmin} ^2$ throughout. This is a good approximation since in all our simulations there are a large number of galaxies in each pixel, so the central limit theorem applies.

\subsection{Forward Modelling the Mask} \label{sec:forward}
\par One advantage of performing inference with full forward models of the data is that we do not need to deconvolve the mask. This is both computationally simpler and avoids the risk of bias from inaccurate deconvolution which is present in the standard likelihood analysis.
\par Given two masked shear fields $a(\theta)$ and $b(\theta)$, a na\"ive estimate of the lensing spectrum is the pixel pseudo-$C_\ell$ spectrum:
\begin{equation}
    \widetilde C^{\text{pix}, EE}_\ell  = \frac{1}{2 \ell + 1} \sum_{m = - \ell} ^{\ell} \langle a^E_{lm} b^E_{lm} \rangle,
\end{equation}
where the tilde is used to denote the fact that we have not corrected for the mask and the `pix' superscript reminds us that we have not accounted for the pixel window function. Analogous expressions are easily found for the $EB$ and $BB$ spectra.
\par In an unmasked field, lensing by large scale structure will only induce power in the $EE$ spectra, but to retain information leaked into the $EB$ and $BB$ spectra due to the presence of a mask, in Pipeline I, we use:
\begin{equation}
   \widetilde C_\ell ^{\text{pix}} =  \widetilde C_\ell ^{\text{pix}, EE} + \widetilde C_\ell ^{\text{pix}, EB} + \widetilde C_\ell ^{\text{pix}, BE} + \widetilde C_\ell ^{\text{pix}, BB},
\end{equation}
as the estimator. This is computed using {\tt HEALpy}~\cite{gorski1999healpix, 2005ApJ...622..759G}. 
\par In a future pipeline it may still be desirable to use the pseudo-$C_\ell$ formalism to avoid mixing between $E$ and $B$-modes, allowing us to immediately remove $B$-modes induced by unknown systematics. As long as the data and theory are treated in the same way the pseudo-$C_\ell$ formalism will not introduce bias, as it could in the standard likelihood analysis.

\begin{figure*}
%\centering
\includegraphics[width=1.\linewidth]{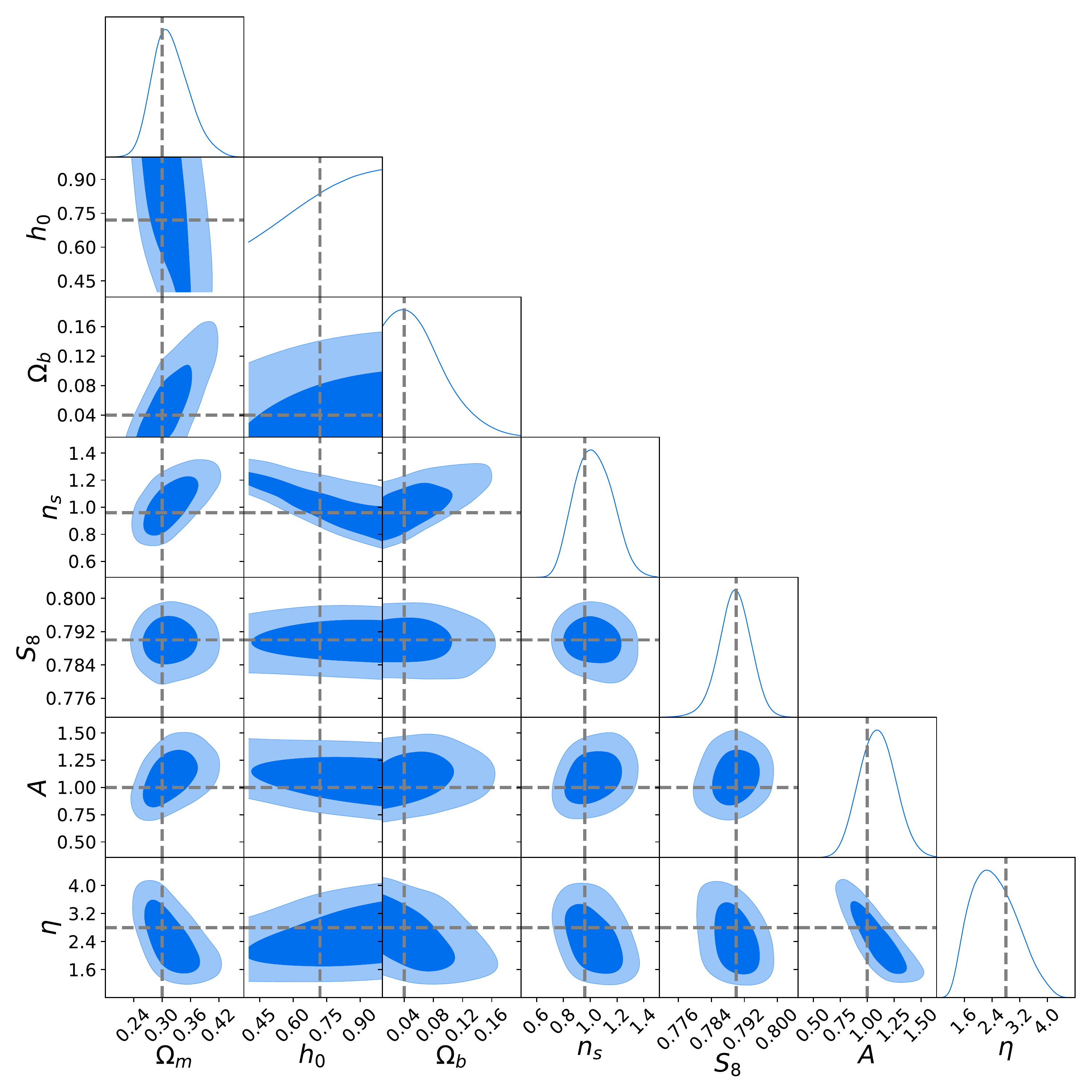}
\caption{$68 \%$ and $95 \%$ credible region parameter constraints found with DELFI analysis DA1 after 1000 simulations, for a mock Stage IV experiment. We confirm that we recover the input cosmology within statistical errors. We plot the convergence in Figure~\ref{fig:train}. In a realistic situation there may be a larger number of nuisance parameters. This would not dramatically slow convergence because we could `nuisance harden' the data compression step, to only learn the posterior for the parameters of interest~\cite{alsingnuissance}.}
\label{fig:fiducial}
\end{figure*}
    
\begin{figure}
\centering
\includegraphics[width = 90mm]{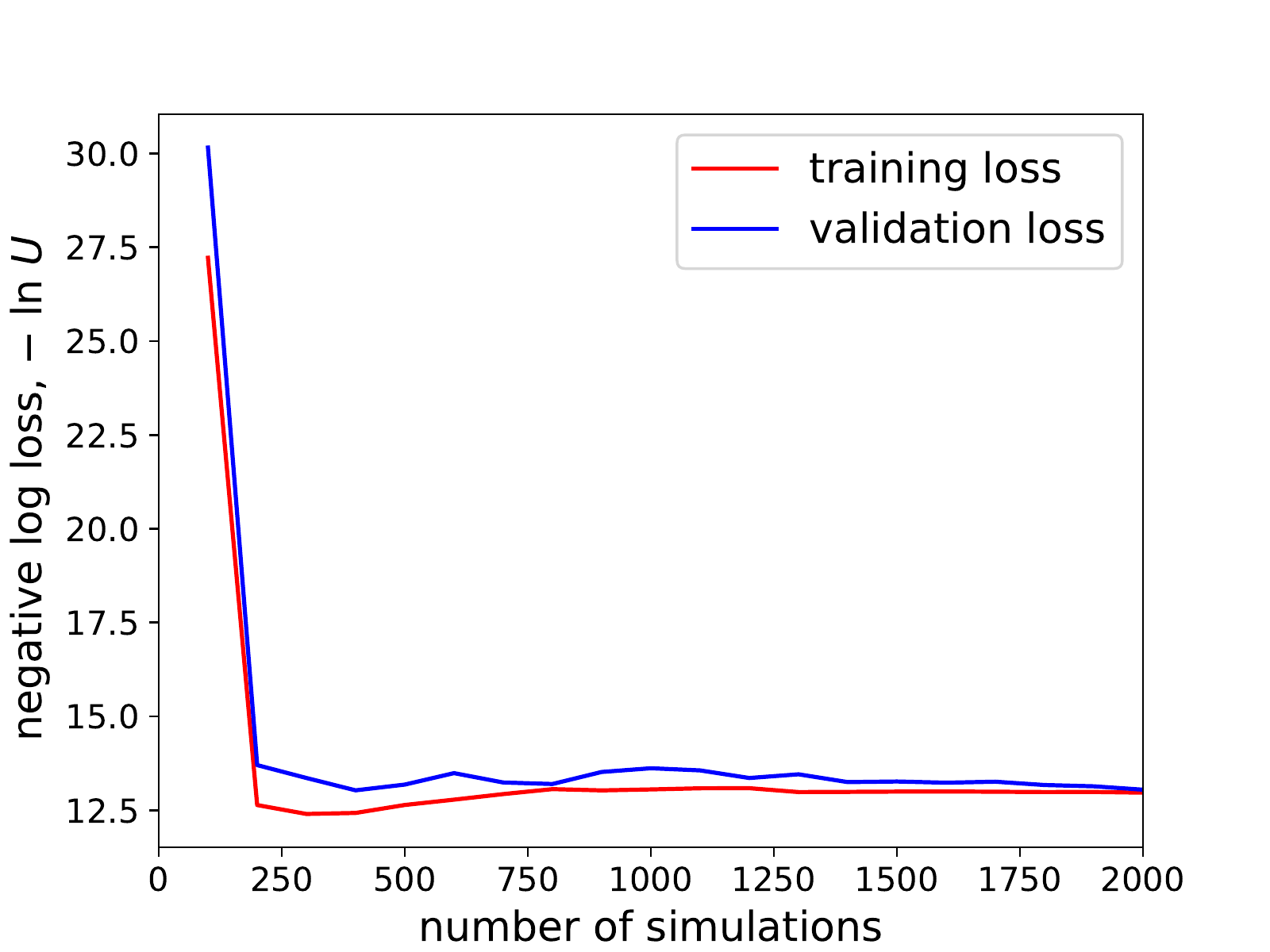}
\caption{The negative loss function defined in equation (\ref{eq:negative loss}) for the training and validation sets as a function of the number of simulations. This suggests that  $\mathcal{O}(1000)$ simulations are needed for a Stage IV experiment. This is similar to the number found in~\cite{alsingndes}, which only considered a simple Gaussian field forward model with no intrinsic alignments, implying that the convergence rate is fairly insensitive to the precise details of the model.}
\label{fig:train}
\end{figure}

\subsection{Mimicking a Stage IV Experiment}
To estimate the number of simulations needed for a Stage IV experiment and ensure that pipeline recovers the input cosmology we produce mock data with Pipeline I. We use 6 tomographic bins sampling 15 logarithmically spaced $\ell$-bins in the range $\ell \in [10,1000].$ We then run {\tt pydelfi} to estimate the posterior distribution of the cosmological parameters for this data. The final parameter constraints for a lambda cold dark matter (LCDM) cosmology with two nuisance intrinsic alignment parameters are shown in Figure~\ref{fig:fiducial}. This confirms that we recover the input parameters within errors. 
\par In Figure~\ref{fig:train} we plot the negative loss function defined in equation (\ref{eq:negative loss}) for the training and validation sets. Both have converged within $\mathcal{O}(1000)$ simulations. This is similar to the number found in the simple Gaussian field pipeline presented in~\cite{alsingndes}, suggesting that the inclusion of higher order effects including intrinsic alignments and non-Gaussian field corrections does not significantly increase the required number of simulations. 
\par When working with real data, we may require a large number of nuisance parameters. Nevertheless, we do not expect this to dramatically increase the number of simulations needed, since we can always tune the data compression to maximize the information retention of the parameters of interest, following the procedure in~\cite{alsingnuissance}. 
\par Each simulation takes approximately 33 minutes on a single thread of a 1.8 GHz {\tt Intel Xeon (E5-2650Lv3) Processor}. Thus if run on 100 threads in parallel, the total simulation time of the DELFI inference step takes only 10 hours. Many of the individual modules in the pipeline are multithreaded (e.g {\tt Flask}), so running on even more threads would further reduce the total run-time.

\section{Prospects for Stage III Data} \label{sec:stage iii}
In this section we discuss the viability of applying analysis DA1 to existing Stage III data. For the remainder of this section we assume a circular mask of $4951 \text{ deg} ^2$, similar to the final coverage of the Dark Energy Survey~\cite{troxel2015intrinsic} with $10 {\rm \ galaxies \ per \ arcmin}^2$ and use Pipeline I throughout this section -- except where modifications are explicitly stated.

\begin{figure}
\centering
\includegraphics[width = 90mm]{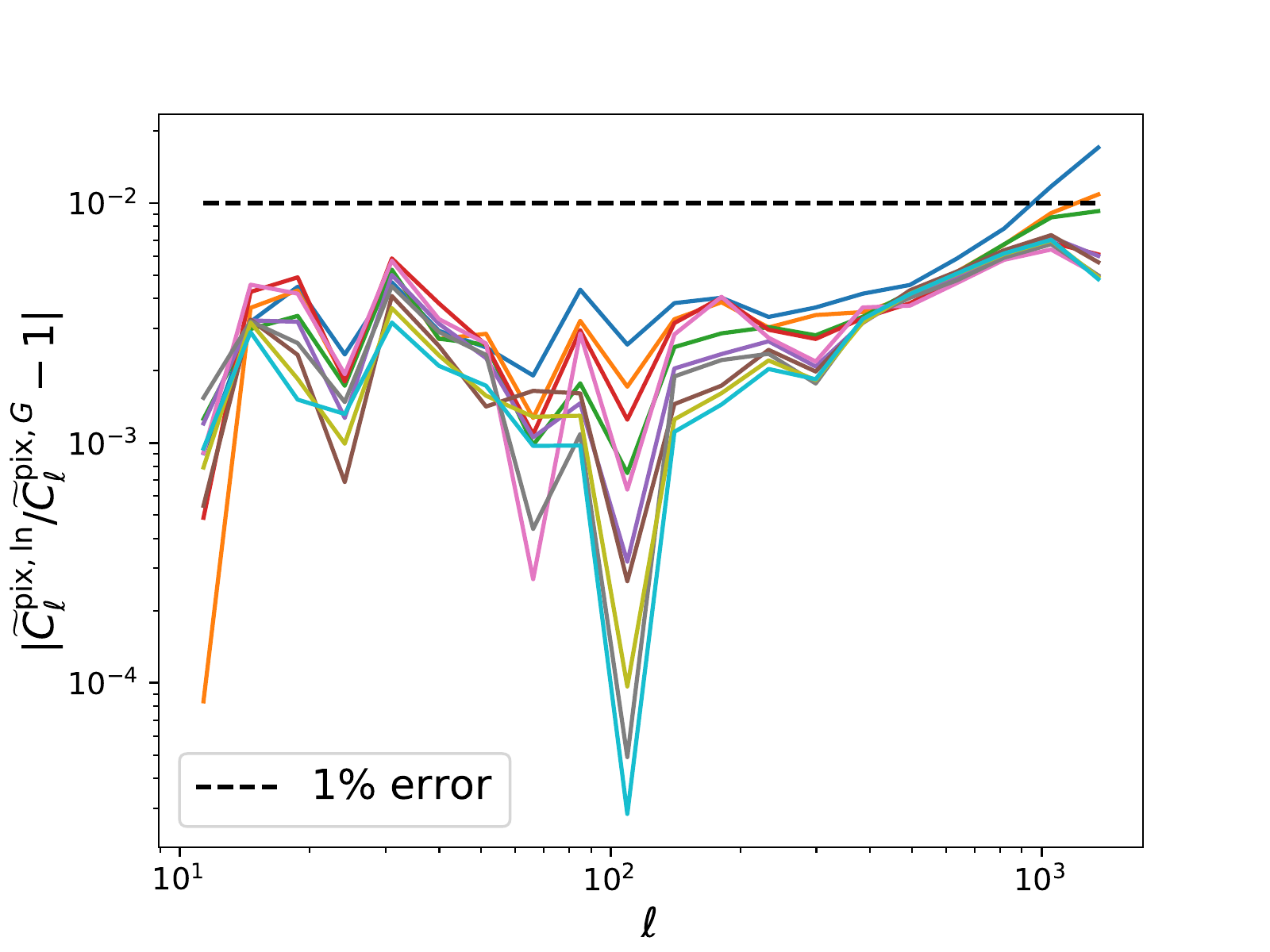}
\caption{The colored lines show the absolute value of the difference between the average recovered cross and intra-bin spectra from 100 lognormal and 100 Gaussian realizations ($4$ tomographic bins, $N_{\text{side}} = 512$ and $\ell \in [10,1535]$). The difference is due to the band-limit bias in the lognormal field discussed in Section~\ref{sec:band-limit}. With these model choices the band-limit bias is safely below $1\%$ for nearly all data points.}
\label{fig:verification}
\end{figure}

\subsection{Validating the Lognormal Simulations}
\par Lognormal fields were used to generate the covariance matrix in the recent Dark Energy Survey Year 1 analysis~\cite{troxel2017dark}. The authors found no difference in parameter constraints between this analysis and one which used a halo model to generate the covariance matrix -- but to verify that our pipeline is ready for Stage III data, we must also ensure that we recover an unbiased $C_{\ell}$ from the maps.
\par Given an accurate input $C_\ell$, the only bias in Pipeline I comes from regularizing the map (see Section~\ref{sec:lognormal}). We would not expect the band-limit bias of the lognormal field to be problematic since imposing a band-limit would affect the data in the same way. However this assumes that the true field is exactly lognormal. Nevertheless we check to ensure that the combined effect of regularization and imposing a band-limit is small. We quantify this statement by finding the difference between the average recovered pixelated $C_\ell$'s from 100 Gaussian simulations (where no band-limit bias or regularization bias is present) and 100 lognormal simulations. Each $4$-tomographic bin simulation takes approximately 15 minutes on a single thread and the difference in the recovered spectra is shown in Figure~\ref{fig:verification}. The bias is safely below $1 \%$ in all but three data points. This confirms that once minor updates have been made (see next subsection), the pipeline will be ready for use on today's data. 

\subsection{Model Improvements}

Only a small number of adjustments must be made to DA1 to apply this analysis to existing data. These are:
\begin{itemize}
\item We must accurately account for baryonic physics. This can be handled using a halo model code~\cite{mead2015accurate}, potentially in combination with the k-cut cosmic shear approach~\cite{taylor2018k, bernardeau2014cosmic}, optimally cutting scales which can not be accurately modeled. 
\item We must introduce several nuisance parameters. As well as allowing for free multiplicative and additive shear biases, photo-z bias parameters will need to be allowed to vary, as in the Dark Energy Year 1 analysis~\cite{troxel2017dark}. This will increase the number of nuisance parameters. To avoid excessive computational costs we must `nuisance harden'~\cite{alsingnuissance} the data compression step.
\end{itemize}

\section{Testing the Gaussian Likelihood Approximation} \label{sec:test gauss like}
In this section we compare DELFI and the standard Gaussian likelihood analysis by running the DA2 analysis and the LA analysis, on the same mock Stage IV data. We use Pipeline II to generate the mock data, produce the covariance matrix and generate the forward models in DA2. Since DELFI does not assume any particular likelihood, differences in the resulting parameter constraints are only due to the Gaussian likelihood assumption in LA. Because we can not just forward model everything in LA, care must be taken to ensure that the band-limit bias, deconvolving the mask, deconvolving the pixel window function and subtracting the shot-noise does not lead to additional bias between the two analyses. Controlling for these effects is described in the first subsection.

\subsection{Modeling Choices In Pipeline II}
To avoid the band-limit bias we use a Gaussian field, rather than the lognormal field.

\par We do not apply a mask in DA2 as we have found that using the pseudo-$C_\ell$ method

(with the public code {\tt NaMaster}~\cite{alonso2018unified}) can bias parameter constraints, with our choice of {\tt HEALpix}~\footnote{\url{https://sourceforge.net/projects/healpix/}}  $N_{\rm side}$ and $\ell_{\rm max}$ by up to $1 \sigma$. Instead we adjust the galaxy number density so that total number of galaxies and hence the signal-to-noise remains unchanged.

\par In LA we decide to take the $C_\ell$, with no shot-noise term in the intra-bin case, as the data vector. Thus we must subtract off the expected value of the noise in DA2. This is computed by running 500 noise-only simulations, as in the analysis of~\cite{hikage2018cosmology}.

\begin{figure}
\centering
\includegraphics[width = 90mm]{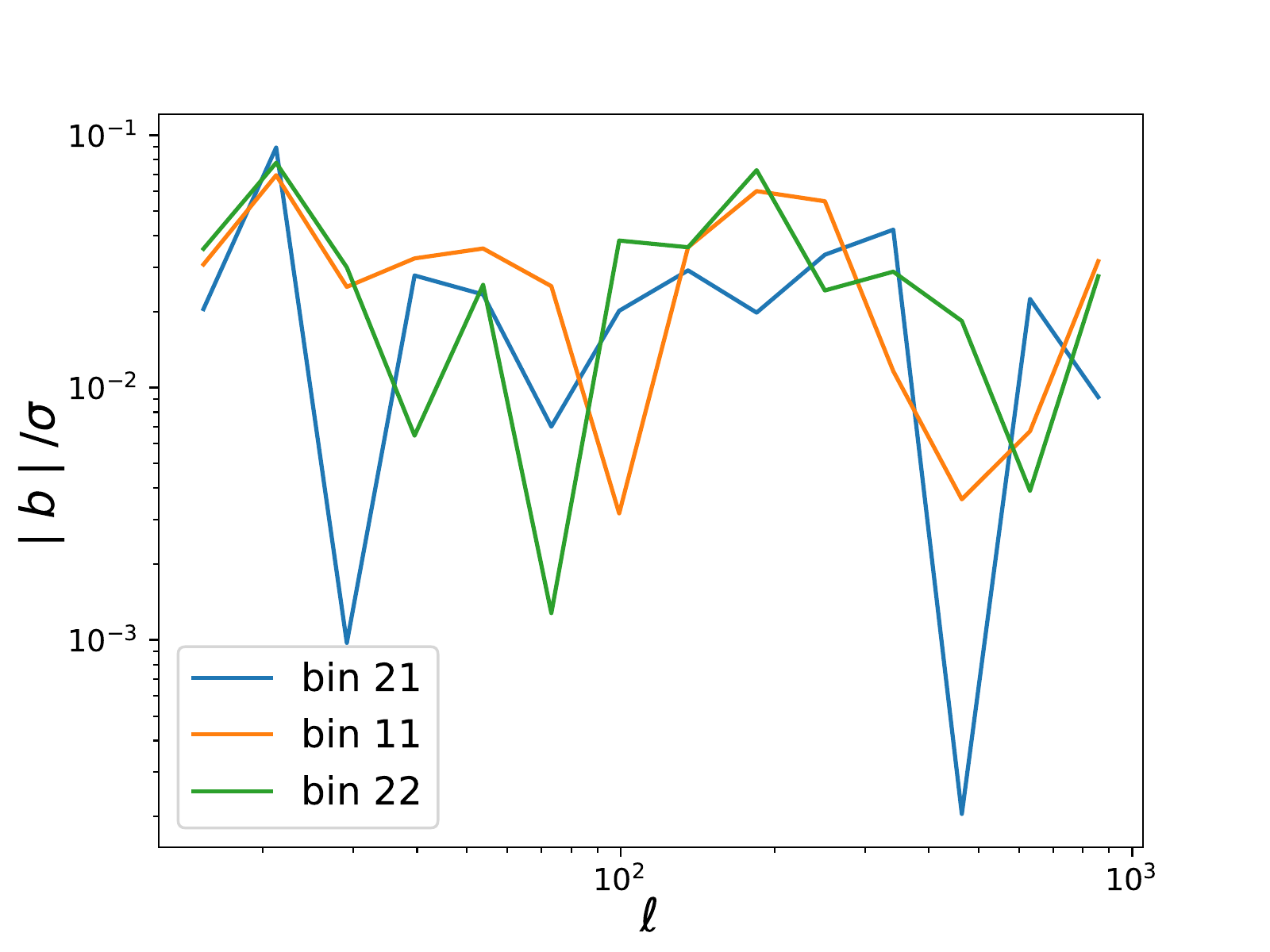}
\caption{The absolute value of the bias, $\mid b \mid$, due to imperfect pixel-window deconvolution and noise subtraction relative to the statistical error, $\sigma$, from 500 Pipeline II simulations. This confirms that the comparison between DELFI and the likelihood analysis presented in Section~\ref{sec:gauss_bias} will be unaffected.}
\label{fig:pipe_II_verify}
\end{figure}
    
We must also account for the fact that the shear spectra are computed on pixelized maps -- that is, we must deconvolve the pixel window function, $w_\ell$, which is defined in~\cite{jeong2014effect} and computed using {\tt HEALpix}. This assumes that the scale of the signal is large relative to the pixel scale and that all pixels are the same shape. The window-corrected spectrum, $ C_\ell$, is given in terms of the spectrum computed from a pixelized map, $ C_\ell^{\text pix}$, by:
\begin{equation}
    C_\ell = w_\ell ^{-2}  C_\ell^{\text pix}.
\end{equation}

\par By running 500 Gaussian field simulations we have confirmed that the combined bias from deconvolving the pixel window function and subtracting the shot-noise is small, so that we can fair comparison between the DA2 analysis and the LA analysis. This is shown in Figure~\ref{fig:pipe_II_verify}. The absolute value of the bias, $|b|$, is small relative to the statistical error, $\sigma$, with $ | b | / \sigma  < 0.1$ for all data points.

\subsection{Impact of the Gaussian Likelihood Approximation} \label{sec:gauss_bias}

\begin{figure}
\centering
\includegraphics[width = 68mm]{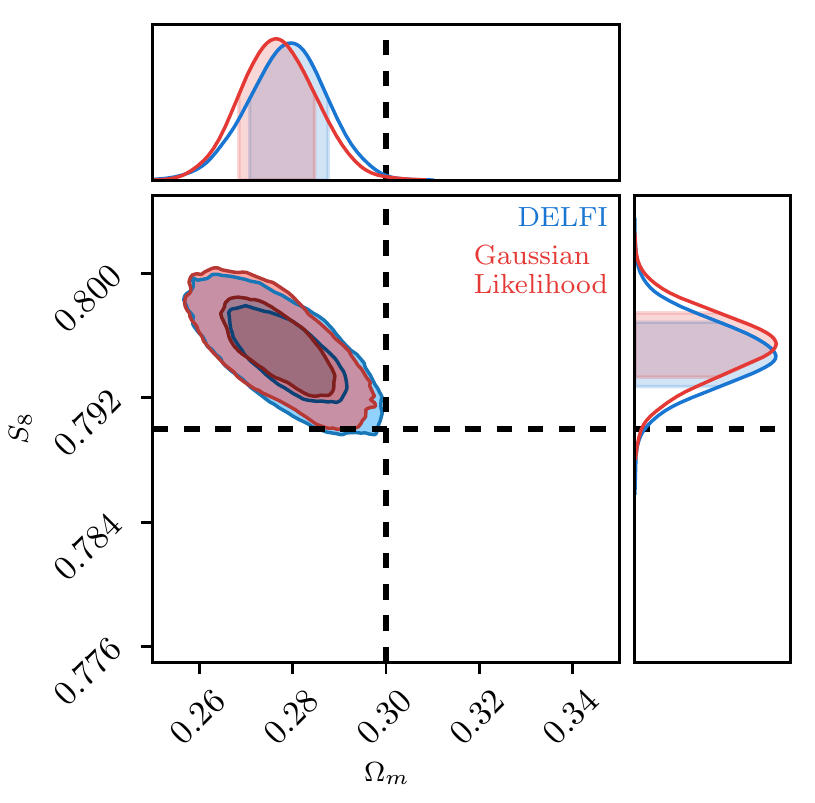}
\includegraphics[width = 68mm]{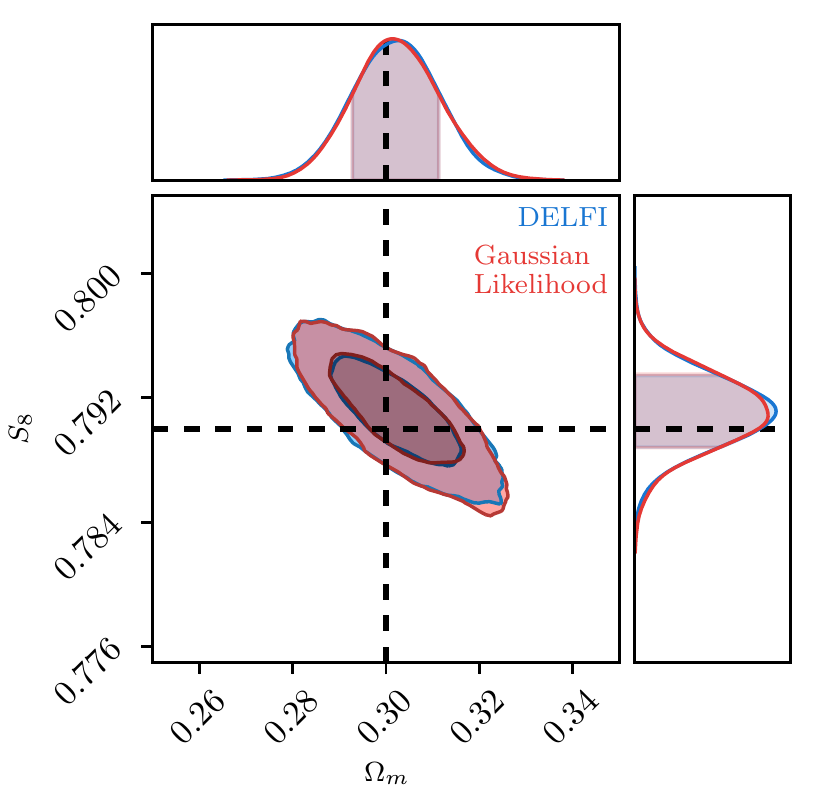}
\includegraphics[width = 68mm]{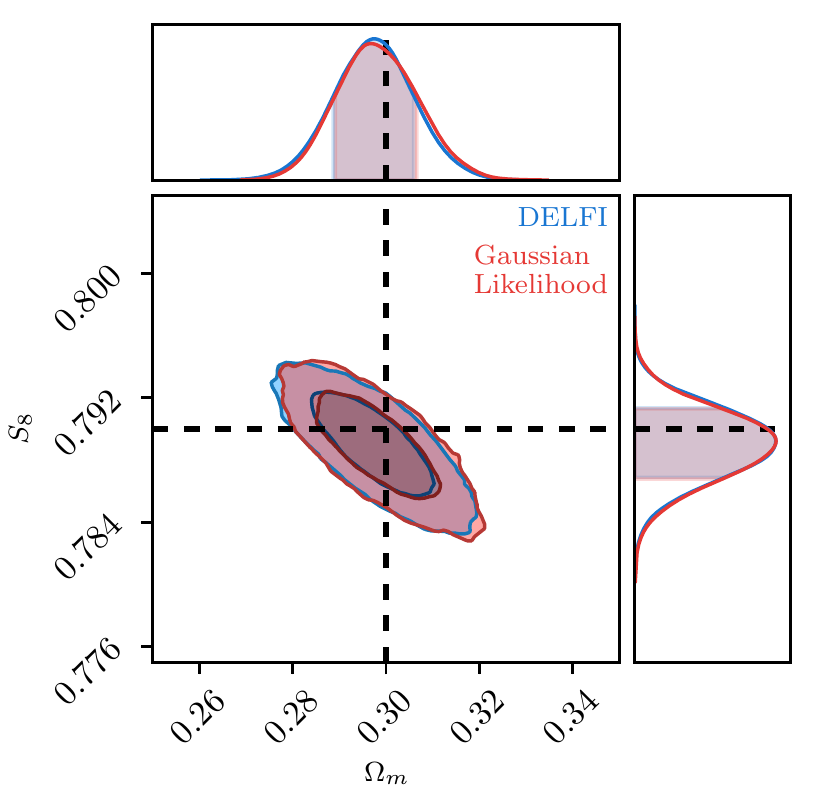}
\caption{The $68 \%$ and $95 \%$ credible region parameter constraints for three random data realizations found using a MCMC Gaussian likelihood analysis and DELFI, which makes no assumption about the functional form of the likelihood. The mock data input cosmology is labeled by black dotted lines. Only in the first realization, does the input cosmology lie outside the $68 \%$ credible region -- but statistically this is to be expected for a small number of realizations. The contours found using the two different analyses are very similar for all three data realizations suggesting that the Gaussian likelihood approximation has negligible impact, and the compression in equation (\ref{eq:compress1}) is lossless. This former statement is confirmed in Figure~\ref{fig:MLE_distro}. }
\label{fig:comparison}
\end{figure}

\begin{figure}
\centering
\includegraphics[width = 80mm]{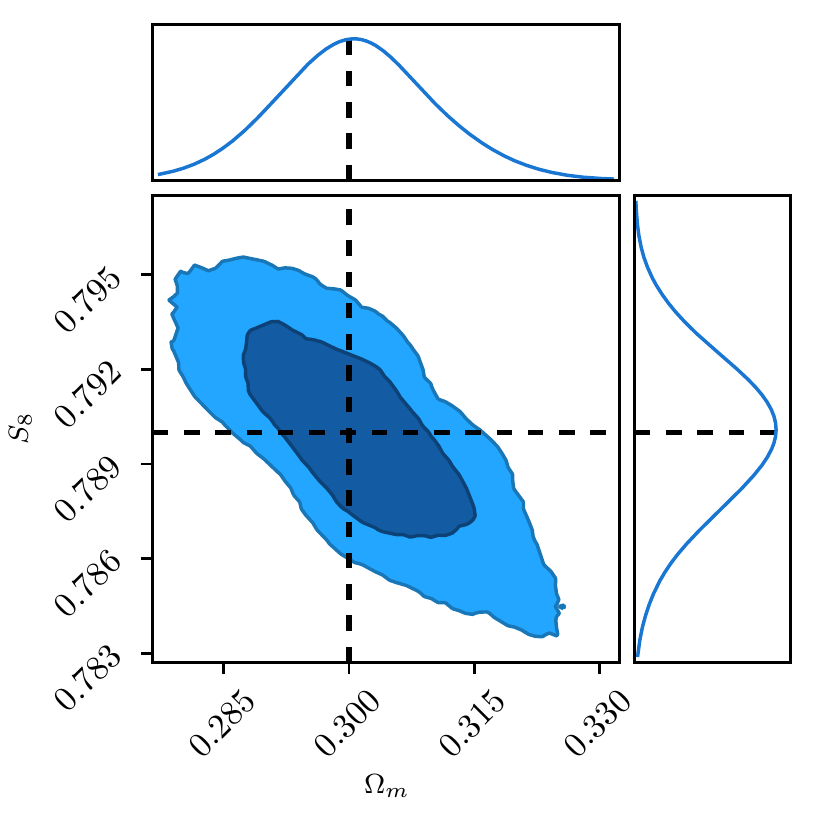}
\caption{The $68 \%$ and $95 \%$ credible region of the MLE distribution, assuming a Gaussian likelihood. The value of the input cosmology is indicated by the black dotted lines, and lies at the center of the contours. This implies that the Gaussian likelihood approximation does not lead to any measurable bias in our setup.}
\label{fig:MLE_distro}
\end{figure}

To test the impact of the Gaussian likelihood approximation we first generate 1000 mock data realizations using Pipeline II. We take 15 logarithmically spaced $\ell$-bins in the range $[10,1000]$ and restrict our attention to the $S_8 - \Omega_m $ plane. To cut computation cost, we use only two tomographic bins. The parameters $S_8$ and $\Omega_m$ primarily impact the amplitude of the shear spectrum, so we do not expect to lose too much information with this choice~\cite{taylornonparametric,spurio2018testing}.
\par It is known from analyses of cosmic microwave background temperature anisotropies that non-Gaussian likelihoods arise even for Gaussian fields~\cite{hamimeche2008likelihood}, as the argument given in Section~\ref{sec:insufficient} holds for any field configuration. Nevertheless we generate $1000$ lognormal realizations in conjunction with the Gaussian fields to determine whether the field configuration impacts the likelihood. The results are plotted in Figure~\ref{fig:hist}. The skew in the likelihood is indistinguishable between the Gaussian and lognormal field configurations, justifying our choice to work with Gaussian fields for the remainder of this section.
\par For three random data realizations, we run a DELFI and a Gaussian likelihood analysis. The resulting posteriors are shown in Figure~\ref{fig:comparison}. Each subplot corresponds to one of the three realizations.
\par In all three cases the DELFI and Gaussian likelihood contours are very similar. This suggests the Gaussian likelihood assumption does not bias parameter constraints in the $S_8 - \Omega_m$ plane and the compression defined in equation (\ref{eq:compress1}) is lossless.
\par To confirm and quantify this statement, we sample the maximum likelihood estimator (MLE) distribution assuming a Gaussian likelihood, using the 1000 data realizations generated earlier. For each realization, the MLE is found using the Nelder-Mead algorithm built into {\tt scipy} and wrapped into {\tt Cosmosis} using the default settings. The resulting MLE distribution is shown in Figure~\ref{fig:MLE_distro}. The input cosmology lies almost exactly at the center of the $68 \%$ credible region which implies that there is no measurable bias from the Gaussian likelihood approximation.
\par We stress that these conclusions only hold for the $C_\ell$ analysis presented in this work. In particular, the  $\ell$-binning strategy matters. By binning $\ell$-modes we are taking a sum over random variables, so by the central limit theorem broader bins correspond to more Gaussian data. This is verified in Figure~\ref{fig:hist} and can be seen by comparing the skew in the second and third row. The Gaussian likelihood approximation could be important for much narrower bins. While the Gaussian likelihood approximation may not be valid for other weak lensing summary statistics, a recent paper has shown it is also valid for two-point correlation functions~\cite{lin2019non}.

\section{Future Prospects} \label{sec:Future }
We review the main known cosmic shear systematics which must eventually be included in the full forward model. To account for many of these effects we must first take the base model presented in this work to `catalog level'. This can be done by first generating a consistent density field -- either with {\tt Flask} or by taking the difference between two neighbouring tomographic bins -- and then populating the density field with a realistic population of galaxies~\cite{miller2007bayesian} assuming a biased tracer model (e.g~\cite{elvin2018dark}). 
Cosmic shear systematics break down into four broad categories: data-processing, theoretical, astrophysical and instrumental systematics. 
\par On the data-processing side, accurately measuring the shape and photometric redshift of galaxies is the primary challenge. Both measurements are dependent on the galaxy-type~\cite{kannawadi2018towards}, and this is in turn correlated with the density through the morphology-density relation~\cite{houghton2015revisiting}. Rather than using the best fit parameters for each galaxy, we can sample the posterior on each galaxy as in a Bayesian hierarchical model~\cite{alsing2015hierarchical} to propagate the measurement uncertainty into the final parameter constraints, as suggested in~\cite{alsingndes}.  We can also account for image blending~\cite{kannawadi2018towards, samuroff2017dark} more easily with forward models.
\par Two important theoretical systematics are the reduced shear correction~\cite{dodelson2006reduced,bartelmann2001weak} and magnification bias~\cite{hamana2001lensing, liu2014impact}. The former correction accounts for the fact that we measure the reduced shear $\gamma / (1- \kappa)$ with a weak lensing experiment. In a likelihood analysis, this can be computed using a perturbative expansion as in ~\cite{dodelson2006reduced, shapiro2009biased}. This is slow and requires us to rely on potentially inaccurate fitting functions for the lensing bispectrum.  Meanwhile the magnification bias accounts for the fact that galaxies of the same luminosity can fall above (below) the detectability limit in regions of high (low) lensing magnification. In both cases, these systematics can be easily handled with full forward models of consistent shear and convergence fields.
\par The two dominant instrumental systematics are the telescope's point spread function (PSF)~\cite{massey2012origins} and the effect of charge transfer inefficiency (CTI) in the charge-coupled devices (CCDs)~\cite{massey2012origins, rhodes2010effects}. Efforts are underway to build pipelines which characterize these effects in upcoming experiments (e.g~\cite{vavrek2016mission} and Paykari et al. (in prep)). Integrating these pipelines into ours would enable the propagation of instrumental errors through to the final parameter constraints.
\par On the astrophysical side, the two dominant systematics are the impact of baryons on the density field~\cite{semboloni2010weak} and the intrinsic alignment of galaxies~\cite{hirata2004intrinsic, kiessling2015galaxy}. For Stage IV data, forward models will likely have to be based on high-resolution N-body lensing simulations~\cite{izard2017ice, kiessling2011sunglass} to include the effects of baryons. Even with today's highest resolution simulations the impact baryons is still uncertain~\cite{huang2018modeling}, so it will likely be necessary to optimally cut~\cite{taylor2018k} or marginalize out uncertain scales~\cite{huang2018modeling}.  Meanwhile more sophisticated intrinsic alignment models which account for different alignment behaviour by galaxy type~\cite{samuroff2018dark} will need to be included. 
\par Eventually higher-order statistics such as peak counts and the shear bispectrum can be added. Since DELFI automatically handles multiple summary statistics in a unified way, the constraints will be tighter than doing the two-point and higher-order statistic analyses separately. With a greater ability to handle systematics, DELFI may also open up the possibility of performing inference with weak lensing flux and size magnification~\cite{alsingndes, alsing2015weak, duncan2013complementarity, huff2013magnificent, hildebrandt2013inferring, heavens2013combining}.

\section{Conclusion} \label{sec:Conslusion}
\par By comparing a Gaussian likelihood analysis to a fully likelihood-free DELFI analysis, we have found that the Gaussian likelihood approximation will have a negligible impact on Stage IV parameters constraints. Nevertheless we recommend the development of DELFI weak lensing pipelines, because they offer the possibility of performing rapid parallel inference on full forward realizations of the shear data. In the future, this will allow us to seamlessly handle astrophysical and detector systematics -- at a minimal computational cost. Since we have shown that applying the standard DELFI data compression (see equation~\ref{eq:compress1}) to the $C_\ell$ summary statistic is lossless (see Figure~\ref{fig:comparison}), this comes at no cost in terms of constraining power. 
\par We have taken the first steps towards developing a pipeline to rapidly generate realistic non-Gaussian shear data, including the impact of intrinsic alignments. These effects are handled in the same way as the Dark Energy Survey  year 1 analysis~\cite{troxel2017dark} and we have verified that the regularisation of the lognormal field will not lead to bias using today's data. Additionally the pipeline is computationally inexpensive, so in the future it will be useful for quickly determining which systematics are important.
\par We confirm the result of~\cite{alsingndes} (which used a simple Gaussian field model for lensing field) that $\mathcal{O} (1000)$ would be required to perform inference on Stage IV data. This suggests that this estimate is robust and largely insensitive precise details of the cosmic shear forward model. 
\par We conclude that DELFI has a promising future in cosmic shear studies.  Developing fast simulations that fully integrate all relevant astrophysical, detector and modelling effects is the primary hurdle. With so many clear advantages over the traditional likelihood analysis, developing these simulations should be a priority.

%\vfil

\section{Acknowledgements} \label{sec:Conslusion}
PLT thanks Luke Pratley, David Alonso and Martin Kilbinger. BDW thanks Chris Hirata for asking him a question a decade ago that is answered in this paper. The authors would like to thank the referee whose comments have helped significantly improve the clarity of this paper. We are indebted to the developers of all public code used in this work. PLT acknowledges the hospitality of the Flatiron Institute. 
\par This work was supported by a collaborative visit funded by the Cosmology and Astroparticle Student and Postdoc Exchange Network (CASPEN). PLT is supported by the UK Science and Technology Facilities Council. TDK is supported by a Royal Society University Research Fellowship. JA was partially supported by the research project grant “Fundamental Physics from Cosmological Surveys” funded by the Swedish Research Council (VR) under Dnr 2017- 04212. BDW is supported by the Simons foundation. The authors acknowledge the support of the Leverhulme Trust.

\bibliographystyle{apsrev4-1.bst}
\bibliography{bibtex.bib}

\end{document}